\documentclass[12pt,preprint]{aastex}
\slugcomment{Accepted and scheduled for publication  
in {\it the Astrophysical Journal},  for  May 20, 2008, v 679 1 issue}  
\def\lax {\ifmmode{_<\atop^{\sim}}\else{${_<\atop^{\sim}}$}\fi}  
\def\gax {\ifmmode{_>\atop^{\sim}}\else{${_>\atop^{\sim}}$}\fi}  
\def\gtorder{\mathrel{\raise.3ex\hbox{$>$}\mkern-14mu
             \lower0.6ex\hbox{$\sim$}}}

\usepackage{lscape}

\begin{document}

\title{On the nature of  the variability power decay towards soft spectral states in X-ray binaries.
 Case study in Cyg X-1}

\author{Lev Titarchuk\altaffilmark{1,2,3} and Nikolai Shaposhnikov\altaffilmark{4}}
%and Vadim Arefiev\altaffilmark{5}  }

\altaffiltext{1}{George Mason University/Center for Earth
Observing and Space Research, Fairfax, VA 22030; and US Naval Research
Laboratory, Code 7655, Washington, DC 20375-5352; ltitarchuk@ssd5.nrl.navy.mil }
\altaffiltext{2}{Dipartimento di Fisica, Universit\'a di Ferrara, via Saragat 1, I--44100, Ferrara, Italy; 
titarchuk@fe.infn.it}
\altaffiltext{3}{Goddard Space Flight Center, NASA, code 661, Greenbelt  
MD 20771; lev@milkyway.gsfc.nasa.gov}
\altaffiltext{4}{Goddard Space Flight Center, NASA/Universities Space Research
Association, code 662, Greenbelt  
MD 20771; nikolai@milkyway.gsfc.nasa.gov}
%\altaffiltext{5}{Space Research Institute (IKI), Russian Academy of Science, Profsoyuznaya 84/32, 117997, Moscow, Russia,
%gita@hea.iki.rssi.ru}

\begin{abstract}

A characteristic feature of the Fourier Power Density Spectrum (PDS) 
observed from black hole X-ray binaries in low/hard and intermediate spectral states
is a broad band-limited noise, 
%roughly 
characterized by a constant below some frequency (a ``break''  frequency) and
a power law above this frequency. It has been shown that the variability  of this 
type can be produced by the inward  diffusion of the local driving perturbations  in a bounded configuration (accretion disk or corona).
%towards the inner region where the energy release occur.
%A perturbation diffusion model successfuly reproduces the form of the observed continuum. 
In the framework of this model, the perturbation diffusion time $t_0$  is related to
the  phenomenological break  frequency, while the PDS power-law slope above the ``break'' is determined
by the viscosity distribution  over the configuration. 
%, recently developed to explain the
%shape of Fourier Power Density Spectrum (PDS), describes
%We show that a perturabation diffusion model of Fourier Power Density Spectrum (PDS) formation 
%in accretion 
%powered X-ray binary systems 
%derived from the first principles of the diffusion theory 
The perturbation diffusion scenario explains the decay of  the power of X-ray variability  
observed in a number of compact  sources (containing black hole and neutron star) during  an evolution of theses sources  from
low/hard to high/soft states.
We compare the model predictions with the subset of data from Cyg X-1 collected by the Rossi  X-ray Time Explorer (RXTE).
%Timing properties of X-ray emission are considered to be a result of diffusive propagation 
%of the driving perturbations in a bounded medium. 
Our extensive analysis of the Cyg X-1 PDSs  demonstrates that 
 the observed integrated  power $P_x$ 
 %well fitted by the model 
 decreases approximately as 
a square root of  the characteristic frequency of the driving oscillations $\nu_{dr}$. 
%We also show that the model integrated power 
%of the resulting PDS, 
%$P_{x,diff}$  is only a small fraction of the integrated power of the driving oscillations $P_{dr}$. 
%In fact,  the model  power $P_{x,diff}$ is related to the variability of mass accretion rate  at the 
%innermost part of  the accretion configuration (geometrically thin disk or Compton cloud).
%   is a result of diffusion propagation of the perturbations   distributed over this configuration (geometrically thin disk or Compton cloud).
%The diffusion theory predicts that   $P_{x,diff}$  is a product  of  $P_{dr}$,  
%$\nu_{dr}^{-1}$ and a reciprocal  of the diffusion time of the perturbation propagation  in  the accretion configuration 
%$t_0^{-1}$.
The RXTE observations of Cyg X-1 allow  us to infer $P_{dr}$ and  $t_0$ as a function of $\nu_{dr}$. 
%We find that $P_{dr}\propto \nu_{dr}^{-1.8\pm 0.16} $and $t_0 \propto \nu_{dr}^{-2.13\pm0.14} $  
Using the inferred dependences of the integrated power of the driving oscillations $P_{dr}$ and $t_0$ on $\nu_{dr}$ we demonstrate that the power predicted by the model 
also decays as  $P_{x,diff} \propto \nu_{dr}^{-0.5}$ that is similar to the observed $P_{x}$ behavior.
% which is likely scaled with the frequency of the local gravity waves inhe disk (Keplerian frequency).  
 %Using a method of  which was developed in one of our previous paper and 
% Our extensive data analysis of the power spectra from Cyg X-1 with an application of the method of Reynolds  ($\rm Re$)  
%number  determination developed in one of our previous paper  indicates  that  ${\rm Re}-$number related to Compton cloud 
%configuration increases when the source evolves toward softer states.   
 %We infer Re number from the observations  
%    We  present r a method to measure an effective Reynolds number, 
%(${\rm Re}$) [inverse of the disk viscosity ($\alpha_{\rm SS})-$parameter]
 We also apply the basic  parameters of observed PDSs, power-law index and  low frequency quasiperiodic 
oscillations, to infer the Reynolds (Re) number from the observations using the method developed in our previous paper.
 Our analysis shows  that  Re$-$number 
 increases from values about  10 in low/hard state to that about  70  during the  high/soft state. 
 \end{abstract}


\section{Introduction}
%\section {Diffusion model}
The main goal of the presented work is to explain a decay of the emergent time variability of X-ray emission in  
compact  sources when these sources evolve from low/hard  (LH) 
to high/soft  (HS) 
states
%likely as a result of the diffusive propagation of the driving  perturbations  [see \cite{tsa07}, hereafter TSA07]   
%evolves and its power steadily decays towards  
%softer state 
[see \citet{rm06} and \citet{tsa07}, hereafter TSA07, for details of observations].  In particular,
power density spectrum (PDS) of black hole binaries in hard states is dominated by a component, which has 
a specific shape roughly described by a broken power-law. The low-frequency part is mostly flat, while 
the power-law  index $\alpha$
above the ``break'' frequency $\nu_{br}$ is variable between 1 and 2.  It is well established that the fractional root-mean-square (rms)
 variability in a source light curve decreases as a source evolves from LH state to HS state. Simultaneously, both $\nu_{br}$ 
 %the break frequency 
 and $\alpha$ 
 %the power index above the break 
 increase. Although empirical shot-noise models were able to
describe in general the observed PDS shape \citep{fws05}, the physical picture explaining the observed evolution during spectral
transitions was missing. Moreover, shot-noise models were challenged by the linear absolute rms-flux relation 
\citep{utt05,utt04}. This rms flux relation assumes that amplitudes and time-scales of shots are not independent, but
are related in some way.
 
%which are distributed over the disk. 
\citet{L97}, hereafter L97,  was the first to suggest a model for this  time variability production in the accretion powered X-ray sources. 
%  detected from many Galactic and extragalactic compact objects (see RM06).  
He  considered small amplitude local fluctuations in the accretion rate at each radius, 
caused by small amplitude variations in the viscosity, and then studied  the effect of these fluctuations on the 
accretion rate at the inner disc edge. His linear calculations show that if the characteristic time-scale of the viscosity 
variations  is everywhere comparable to the viscous (inflow) time-scale, and 
if the amplitude of the variations is independent of radius, then the power spectrum of luminosity fluctuations is a power-law 
$1/\nu$.  If the amplitude of the variations increases with radius, the slope of the power spectrum of the luminosity variations 
is steeper than 1. Lyubarskii pointed out that he had no physical model for the cause of such fluctuations. 
\citet{utt05} pointed out that  rms-flux relation is naturally explained in the framework developed by Lyubarskii.

TSA07 formulated and solved the problem of local driving perturbation diffusion
in a ``disk-like'' configuration (which can be either a geometric thin Keplerian accretion disk or Compton cloud).  The problem of the diffusive propagation of the  space distributed  high-frequency 
perturbations is formulated  as a  problem in terms of  the diffusion equation for  the surface density perturbations. 
This equation is combined with the appropriate boundary conditions.  
The formulation of this problem and its solution are  general and  classical. 
The parameters of the resulting PDS,  diffusion time scale of the diffusion propagation of the 
local perturbations $t_0$ and the power-law index of the viscosity distribution over radius,  
are essential parameters of  diffusion in a given  bounded configuration. 
In TSA07 we call our PDS model for the  Green's function of the bounded configuration as
%band-limited noise component 
a white-red noise (WRN) and we adopt  this 
name throughout this paper. 

The problem formulation was similar to the Lyubarskii's scheme. 
However, the method of solution is different. Lyubarskii treated the factorization of the 
driving term (i.e. separating it into two parts each depending on time and radius only) by 
linearizing the system. In TSA07 the analytical solution  is obtained for  the case of the factorized driving sources. Then,
using the mean value theorem  we showed that the general solution is simply a convolution of the configuration response  signal (the Green's function)   and  the mean driving signal in the configuration.
Thus the resulting power spectrum of the X-ray signal, as a  convolution, is a product 
of the power spectrum related to the  configuration Green's function  and that  related 
to  the perturbation sources (sources of driving oscillations). 

The PDS of the  Green's function is a white-red noise power spectrum (WRN). 
 Specifically, the low frequency (LF) asymptotic form of the WRN PDS, when the frequency  is less than the inverse of   diffusion timescale in the ``disk-like'' configuration  $t_0^{-1}$,
 %[Eqs. (\ref{g_pw_sp_lowfr}), (\ref{g_pw_sp_highfr})] 
 is characterized by a flat shoulder (white noise). 
 In other words, the  LF white noise shoulder  is insensitive to the source and  viscosity distributions over radius.  
 %in the disk 
% as a function of radius. 
 The high frequency (HF)  asymptotic form of WRN is a power law $\nu^{-\alpha}$ with  
 index  $\alpha$, which is determined by the viscosity and perturbation source
 distributions over the accretion configuration. The index $\alpha=3/2$ when the viscosity {\it linearly} increases with radius and  the perturbation sources distribution  is quasi-uniform. 
The basis of the presented power spectrum formation scenario is that  the timing signal of the WRN PDS shape  is a result of diffusive propagation of driving  perturbations in the bounded configuration (disk or Compton cloud)   in the same way as X-ray photon spectrum is a result of the photon diffusion  (namely, upscattering of seed photons) in the  same bounded configuration.

The driving oscillation amplitude is assumed to be a smooth function of the radius.
TSA07 suggested that driving fluctuations in the configuration 
can be introduced by g-mode driving oscillations  at any given  annulus.  
The local g-mode  driving  fluctuations,  produced possibly by  local  Rayleigh-Taylor local instabilities,
%the dynamo as a small-scale stochastic phenomenon,  operating on roughly the local dynamical time-scale,  
are  high-frequency damped quasi-periodic oscillations (QPOs)  which frequencies are related to the local Keplerian 
frequencies. As we mentioned above TSA07 formulated and solved 
a problem of the diffusive propagation of the  space distributed  high-frequency 
perturbations in the bounded configuration.
Our diffusion model for PDS is a product of WRN PDS and the driving source PDS (Lorentzian). 
%is formulated  as a diffusion problem.
% in terms of  the diffusion equation for  the surface density perturbations. 
%This equation is combined with the appropriate boundary conditions.  
%The formulation of this problem and its solution are  general and  classical. 
%The parameters of the resulting PDS,  diffusion time scale of the diffusion propagation of the 
%local perturbations and the power-law index of the viscosity distribution over the disk-like configuration,  
%are essential parameters of  diffusion in a given  bounded configuration. 
%In TSA07 we call our model for band-limited noise component a white-red noise (WRN) and we adopt  this 
%name througout this paper. 

% They also  show that the emergent power density spectra (PDS) as a result of the perturbation  diffusion 
%(white-red noise type of PDS)   is a
%particular signature  of any bounded disk-like configuration.

%solved  a problem  of the diffusive propagation of the driving  perturbations). 
%We demonstrate that the solution of this problem is reduced to the solution of the initial value problem with distributed sources at the initial moment. 

% of the distributed seed photons.  
 %(see sections 2, 3).

%One can speculate  that our solution  is a combination of  independent, uncorrelated shots and thus  such  shot 
%noise model should be  been ruled out by the observed rms-flux relations and lognormal flux distributions in the 
%data (e.g. Uttley, McHardy \& Vaughan 2005).  

% convolution of the solution of the initial value problem and the source distribution function. 
%(see  \S 2.1 and \S 3 ). 
The WRN PDS is  a power spectrum of the solution of the initial value (Cauchy) problem which
%, as well-known, 
is   a linear superposition of exponential  shots [see \citet{{wo01}}].
%which are {\it not independent}. 
For example, if the driving perturbations  are distributed  according to the first eigen-function of the diffusion operator 
%(see \S 3.4) 
then the bounded medium works as a filter producing just one exponential shot as a result of the diffusive  propagation of eigen-function distribution of the seed  perturbations. In the general case the resulting signal is a linear 
superposition of exponential shots which are  {\it related} to the appropriate eigen-functions. 
Furthermore,  
TSA07 demonstrate that  {\it the observed rms-flux relations} [e.g. \citet{utt05}]
%Uttley, McHardy \& Vaughan 2005)
 is naturally explained by our diffusion model.  In the framework of the linear diffusion theory  the emergent 
perturbations are always linearly related to the driving source perturbations through a convolution of the configuration Green's  function and  source distribution.
% (see Appendix B and \S 2.1).

  %distributed  over the disk [$f(R)$ and its derivatives are bounded] 
  
%  We remind a reader that the resulting power spectrum is a product of the white-red noise (WRN) and driving oscillation power spectra (see formulas \ref{pwsp_varphi_res} and
%\ref{pwsp_product}).   
An important  question is  what our diffusion model predicts for relative contributions of the WRN PDS,  the driving  oscillation PDS in the resulting PDS and for a dependence of the integrated PDS power 
%$P_{x, diff}$ 
on  the driving oscillation frequency. 
%$\nu_{dr}$.   
The next question is   how this model dependence of the integrated power 
%$P_{x,diff}$  
vs the driving oscillation frequency 
%$\nu_{dr}$ 
is related to the observed dependence of that.
%$P_x$ vs $\nu_{dr}$.  
The answers to these questions  are the points   of the presented study. 

In \S 2 we refer to   details  of Cyg X-1 observations with RXTE. 
%(see a more detailed description  in TSA07). 
In \S 3 we outline the main features of the 
diffusion model and related formulas. In \S 4 we show how the model integrated power vs  
the driving oscillation frequency
%$P_{x,diff}$ vs $\nu_{dr}$  using the diffusion model  
 fits X-ray data from Cyg X-1.  In \S 5 we present the inferred correlation of the Reynolds  number
with   the driving oscillation frequency 
%$\nu_{dr}$  
and the spectral state (photon index). Application of the paper results to the observed index-QPO frequency correlations is considered in \S 6.
Discussion and final conclusions follow in \S 7. 
 
\section{Observations}

For our analysis we used Cyg X-1   data from the Proportional Counter Array (PCA) and
All-Sky  Monitor (ASM) onboard {\it RXTE} [\citet{swank}]. 
%and  the medium energy (ME) detector of EXOSAT satellite 
%(Turner, Smith \& Zimmermann 1981). 
The data are available through 
the GSFC public archive \footnote{http://heasarc.gsfc.nasa.gov}. In this Paper we
present the analysis of a representative subset of {\it RXTE} observations of Cyg X-1.
 A reader can find the details of data reduction and analysis  
 in \citet{st06} and TSA07. We chose approximately 200 observations to cover the
complete dynamical range of the source evolution from low/hard to high/soft state.
For the presented analysis we refit PDSs with our new model and we used the 
results of our previous spectral analysis for photon index $\Gamma$.

To fit a PDS we used a sum of our perturbation diffusion model and one or two
Lorentzians to  account for the Low Frequency Quasi-Periodic Oscillations (LFQPOs).
For higher photon inicies, when the source is close to high/soft state, 
the contribution of the accretion disk variability component sometimes becomes
significant. It is observed as an additional power law at the lower frequencies (see TSA07 for details).
We fit this component with simple power law, when it is needed.   

%These data cover the period 1996 - 2006 (MJD range $\sim$ 50100 - 53800). 

\section{The main features of the model and underlying assumptions}
%In this section  we remind the reader the fact  of the power spectrum formation in the disk-like configuration which can be either the standard geometrically thin Keplerian disk or quasi-spherical  Compton Cloud (see more details in TSA07).  
%In this section   the fact  of the power spectrum formation in the disk-like configuration
We consider  a scenario related to our model (see also TSA07) where 
%two  ``disk-like'' accretion configurations 
the Compton Cloud ( ``disk-like'' configuration) is located in the innermost part of the source  and the  Keplerian disk is extended from  the Compton Cloud (CC) to the optical companion
[see Fig. 1 in \cite{tf04}, hereafter TF04,  for the model geometry).

We remind the reader that the diffusion equation for the surface density  perturbation is the same 
for any ``disk-like'' configuration for which   the rotational frequency $\Omega(R)$ as a function of radius $R$ has a Keplerian-like profile namely $\Omega\propto R^{-3/2}$ [see \cite{wo01} for details].
%given that it is derived for of, namely $\Omega\propto R^{-3/2}$. 
The standard Keplerian disk  and  ADAF kind of flow (Compton Corona) are particular examples of these accretion ``disk-like''' configurations. 

%We consider  a scenario related to our model (see also TSA07) where 
%two  ``disk-like'' accretion configurations 
%the Compton Cloud (``disk-like'' configuration) is located in the innermost part of the source  and the  Keplerian disk is extended from  the Compton Cloud (CC) to the optical companion
%[see Fig. 1 in \cite{tf04}, hereafter TF04,  for the model geometry).
 
  The CC  presumably contracts when the source goes to the softer state.  TF04  infer  that 
the CC cloud in low/hard state is about 40 Schzwarzchild radii ($R_{\rm S}$)  whereas that in high/soft state is about 4-5 $R_{\rm S}$.
It is obvious that the variability of the  mass accretion rate in  the cloud leads  to the variability of  the gravitational energy release there.   

The Earth observer sees this  mass accretion rate ($\dot M$) variation  as a variability of X-ray flux coming from the source given that   $\dot M$ regulates the supply of the soft seed photons up-scattered off CC hot electrons.   According to our scenario  in  the equatorial plane the plasma is  more dense and consequently colder than that in the CC outer parts.  The soft photons are produced in these relatively cold equatorial layers of CC   and then they are Comptonized off hot electrons of the CC outer part  forming the resulting  X-ray spectrum.  

Thus  one can conclude that the variability of X-ray signal at energies higher than 3 keV (RXTE energy range)   is mostly  determined by   the  fluctuations of the  luminosity $\Delta L_x(t)$  originated  in the CC.  Here we study   the luminosity fluctuations $\Delta L_x(t)$  in the CC configuration.

%In this paper {\it we study the diffusion of the surface density perturbations in the CC configuration}. 

 We assume that  the mass accretion rate variations $\Delta \dot M(0,t)$ is converted with efficiency $\varepsilon_{eff}$ into 
the variations of the X-ray luminosity, i.e. $\Delta L_x(t)= \varepsilon_{eff}\Delta \dot M(0,t)$.
TSA07 show that the  fluctuations of 
%the  luminosity is 
the resulting X-ray oscillation signal $\Delta L_x(t)$ due to the diffusion of the driving perturbations is
 \begin{equation}
\Delta L_x(t)=\int_0^t \varphi(t^{\prime}) Y(t-t^{\prime})dt^{\prime},
\label{convdL}
\end{equation}
i.e. a convolution of the Green's function of the bounded configuration $Y(t)$ (WRN) and 
the source variability function $\varphi(t)$. 
The resulting power spectrum is 
\begin{equation}
||F_{x}(\omega)||^2=||F_{\varphi}(\omega)||^2||F_{Y}(\omega)||^2
\label{pwsp_varphi_res}
\end{equation}
where $F_{x}(\omega),~F_{\varphi}(\omega),~F_{Y}(\omega)$ are  Fourier transforms of 
$\Delta L_x(t), ~\varphi(t),~Y(t)$ respectively.
The WRN PDS  $||F_{Y}(\omega)||^2$   is described  by Eq. (64) in TSA07  (also see the asymptotic forms of that  in Eqs. 65-66  there).
% [see e.g. Eq.  7  in TSA07 for definition of the Fourier transform].

 Using the  total power of the driving oscillations $P_{dr}$ 
 %(see Eq. \ref{total_mod1})
 %$P_{dr}=0.45~{\rm rms}^2$
 %the inferred total power $P_{tot}=0.45~{\rm rms}^2$ with an assumption that $P_{dr}\sim P_{tot}$ 
 one can  present   the driving oscillation PDS as (see Eqs. 22 and B5 in TSA07) 
 \begin{equation}
||F_{\varphi}(\nu)||_{\nu}^2=\frac{\hat\Gamma_{dr}  P_{dr}/(a{\pi})}{(\nu-\nu_{dr})^2 + (\hat\Gamma_{dr}/2)^{2}}
 \label{driv_norm}
 \end{equation}
where $\nu_{dr}=\omega_{dr}/(2\pi)$ is the driving oscillation frequency,  $\hat \Gamma_{dr}$ is a full width of half maximum (FWHM) of the 
Lorentzian and a constant $a$ varies in the range between 
$1$ and $2$ depending on the ratio of $2\nu_{dr}/\hat\Gamma_{dr}$:
\begin{equation}
a=1+\frac{2\arctan(2\nu_{dr}/\hat\Gamma_{dr})}{\pi}.
\label{par_a}
\end{equation}
 For example $a=1$ and $a=2$ when 
$2\nu_{dr}/\hat\Gamma_{dr}\ll 1$ and $2\nu_{dr}/\hat\Gamma_{dr}\gg1$ respectively. 

It is worth noting that Eqs (\ref{convdL}-\ref{par_a}) have been derived in TSA07 with an assumption that
 the local  driving perturbations $\Phi(t,\xi)$  are  damped quasiperiodic oscillations, namely
\begin{equation}
\Phi(t,\xi)=A_{\Phi}f(\xi)\exp(-\frac{1}{2}\Gamma_{dr} t+i\omega_{dr}t)
\label{driving_fourier}
\end{equation}
where the amplitude of the driving oscillation is function of $R$ only ($\xi=R^{1/2}$).
Note the damping coefficients $\hat\Gamma$  in Eq. (\ref{driv_norm}) and $\Gamma$   in Eq. (\ref{driving_fourier})  are related as $\Gamma =2\pi\hat\Gamma$.

Then TSA07  apply the mean-value theorem for the integral $W(R,t)$  of Eq. B2 in TSA07.
Namely the mean-value  theorem in its general form for the product of two continuos  functions $g(t)$ and $f(t)$ states that 
if a  function $g(t)$ is positive, i.e. $g(t)>0$ then 
$$
\int_a^bg(t)f(t)dt = f(x)\int_a^bg(t)dt
$$ 
where $a\leq x\leq b$. 
Given that  the product of the Green's function $G(R,\xi, t)$ and the driving oscillation amplitude  $f(\xi)$ in Eq. B2 in TSA07 is positive  
we obtain that 
$$
W(R,t)= 
A_{\Phi}\int_0^t dt^{\prime} \exp[-\Gamma_{dr}t^{\prime}/2+i\omega_{dr }(\xi_*)t^{\prime}] \int_{R_{in}}^{R_0}G(R,\xi, t-t^{\prime}) f(\xi)d\xi 
$$
where $R_{in}\leq\xi_{\ast}\leq R_0$. Then TSA07  proceed  with the Fourier transformation and PDS determination of $W(R,t)$ and ultimately they derive equations (\ref{convdL}-\ref{par_a}).

\begin{figure}[ptbptbptb]
\includegraphics[width=5in,height=7.1in,angle=-90]{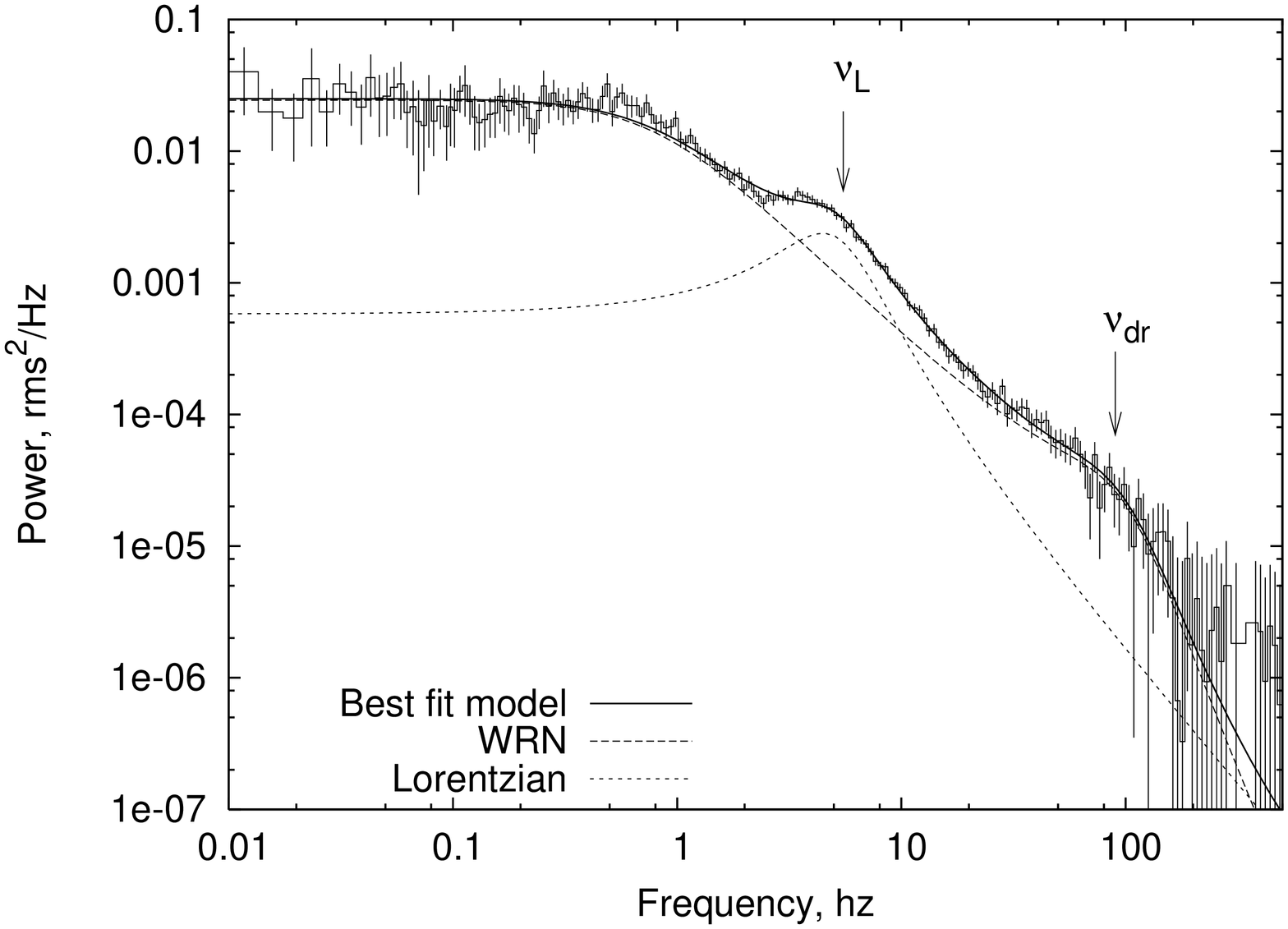}
\caption{A particular example of observable PDS. The PDS continuum is fitted by our diffusion PDS model  which is   a product of WRN PDS and  the driving oscillation Lorentzian. 
% The model PDS  gives  $P(\nu)\propto \nu^{-\alpha}$ on the low frequency side of  $\nu_{dr}$ and $P(\nu)\propto \nu^{-\alpha-2}$ on the high frequency side.  
We also use  a simple Lorentzian to fit QPO features. 
 }
\label{pds_model}
\end{figure}

In Fig. \ref{pds_model} we show a typical example of the model fit to the data using 
Eq. (\ref{pwsp_varphi_res}). The parameters of PDS continuum for white-red noise component 
(WRN) $||F_{Y}(\omega)||^2$ (see for details TSA07) are  the diffusion time scale $t_0$,  index  of the power-law 
viscosity distribution $\psi$ and for the driving oscillation component $||F_{\varphi}(\omega)||^2$ they are 
$\nu_{dr}$ and $\hat\Gamma_{dr}$  
(see Eq. \ref{driv_norm}).  
% One or two relatively broad Lorentzians  are also needed for fitting of QPO features 
%observed in the low-hard  and  intermediate states of Cyg X-1. 
%(see details in TSA07 and above) 

% Note that here the 
%observational PDS is presented  in units of rms$^2$/Hz.  

% throughout the paper. 

%The model parameters of the PDS continuum are  
The power-law index of the viscosity distribution 
$\psi$ is related to the power-law  index of the red noise in the WRN PDS (see TSA07):
$$
\alpha=\frac{3}{2}-\delta=\frac{16-5\psi}{2(4-\psi)} ~~~~ {\rm for~~~\psi>0} 
$$
%\label{in}
%\end{equation}
and 
\begin{equation}
\alpha=2 ~~~~ {\rm for~~~\psi<0}.
\label{index_pds}
\end{equation}
%the diffusion time scale  in the disk-like configuration $t_0$,  $\hat \Gamma_{dr}$ and $\nu_{dr}$.

The model predicted integrated power is (see TSA07)
\begin{equation}
P_{x,diff}\sim\frac{P_{dr}}{2\pi\nu_{dr}t_0({\cal Q}+1/4{\cal Q})DC}
%=\pi a \frac{\nu_{dr}^2 + (\hat\Gamma_{dr}/2)^{2}}{\hat\Gamma_{dr} \nu_{dr}t_0{\cal Q}D}\frac{||F_{x}(0)||^2}{||F_{Y}(0))||^2}.
\label{theory_obs_form}
\end{equation}
where ${\cal Q}=\hat\Gamma_{dr}/\nu_{dr}$ is a quality factor for the driving signal and $D$ is a factor of order of unity.
Equation  (\ref{theory_obs_form}) was derived use the mean value theorem for the integral of the product of two 
functions. In TSA07 we assumed that  a constant $C$  related to  the mean value of  
$||F_{\varphi}(\nu)||_{\nu}^2$  over the frequency integration range is about a few (see Appendix B2 in TSA07 for details). 
Here we specify and obtain $C-$constant  when we compare the  model dependence $P_{x,diff}$ on $\nu_{dr}$ with the observable  
$P_{x}$ on $\nu_{dr}$.  In order to make this comparison one should determine the best-fit parameter $t_0$ and $P_{dr}$ as 
functions of 
%the best-fit 
$\nu_{dr}$. Note that 
to derive Eq. (\ref{theory_obs_form}) we also use  the fact that the  WRN PDS $||F_{Y}(\omega)||^2$  is  
normalized to $1/(Dt_0)$ where $D\gax1$. (see Eq. B16 in TSA07).

Now we follow the method suggested in TSA07 to infer  $P_{dr}$ vs $\nu_{dr}$ from the observations. 
Namely, given the fact that  the driving PDS is a constant at frequencies  $\nu\ll\nu_{dr}$  we have 
 \begin{equation}
||F_{\varphi}(\nu)||_{\nu}^2=|| F_{\varphi}(0)||_{\nu}^2=\frac{\hat\Gamma_{dr}  P_{dr}/(a\pi)}{\nu_{dr}^2 + (\hat\Gamma_{dr}/2)^{2}}.
%\frac{\nu_{dr} P_{tot}/(2\pi Q)}{\nu_{dr}^2 + (\nu_{dr}/2Q)^{2}}.
 \label{driv_norm0}
 \end{equation}
Because for any power spectrum $||F(\omega)||^2$ 
$$|| F(\omega)||^2d\omega=|| F(2\pi\nu)||^22\pi d\nu=||F(\nu)||_{\nu}^2d{\nu}$$ 
we obtain that (compare with Eq. \ref{pwsp_varphi_res})
\begin{equation}
|| F_x(\nu)||_{\nu}^2=(2\pi)^{-1} ||F_{\varphi}(\nu)||_{\nu}^2||F_{Y}(\nu)||_{\nu}^2.
 \label{Fnu_product}
 \end{equation}

Thus a combination of  Eqs. (\ref{driv_norm0}) and  (\ref{Fnu_product})
leads us to  determination of   the integrated power of the driving oscillations 
%$P_{dr}$
 \begin{equation}
P_{dr} = \frac{a\pi(\nu_{dr}^2 + (\hat\Gamma_{dr}/2)^{2})}{\hat\Gamma_{dr} }||F_{\varphi}(0)||_{\nu}^2=
2\pi^2a \frac{\nu_{dr}^2 + (\hat\Gamma_{dr}/2)^{2}}{\hat\Gamma_{dr} }\frac{||F_{x}(0)||^2}{||F_{Y}(0))||^2}.
%=\frac{\hat\Gamma_{dr}  P_{dr}/(a\pi)}{\nu_{dr}^2 + (\hat\Gamma_{dr}/2)^{2}}.
%\frac{\nu_{dr} P_{tot}/(2\pi Q)}{\nu_{dr}^2 + (\nu_{dr}/2Q)^{2}}.
 \label{driv_norm1}
 \end{equation}
We remind a reader that the values of $\nu_{dr}$ and $\hat\Gamma_{dr}$ are the best-fit PDS parameters, $||F_{x}(0)||^2$ is the 
observed PDS value at  $\nu=0$ and  $||F_{Y}(0))||^2$ is  a value of the normalized WRN PDS at $\nu=0$.  As we mention above  the integral of the normalized WRN PDS  $||F_{Y}(\nu))||^2$  over 
%frequency 
$\nu$ is   $1/(Dt_0)$.

\section{The integrated power vs driving oscillation frequency and photon index} 
In Figure \ref{nu_dr_vs_gamma} we present the observed correlation of low 
frequency QPO centroid $\nu_{L}$ with the driving QPO frequency $\nu_{dr}$ (upper panel) 
and photon index $\Gamma$ with $\nu_{dr}$ (lower panel).
These correlations imply  that   $\nu_{L}$ along with $\nu_{dr}$ increase when the source becomes softer. 
In other words, the emission area [Compton cloud (CC)]  contracts when the source evolves to the soft states.  

In Figure \ref{t0_vs_vdr} we also see this  effect of CC contraction as anticorrelation of the CC diffusion 
time scale $t_0$ with  $\nu_{dr}$. The inferred dependence of $t_0$ vs $\nu_{dr}$ can be fitted 
by the power law $t_0 \propto \nu_{dr}^{-2.13\pm0.14}$.

We  infer the integrated power of the driving oscillations $P_{dr}$  vs  $\nu_{dr}$ 
(see Eq. \ref{driv_norm1})
%using $\nu_{dr}$, $\hat\Gamma_{dr}$, $t_0$ and 
and then we obtain the  model integrated power  $P_{x, diff}$ vs $\nu_{dr}$  (see Eq. \ref{theory_obs_form}). 
In Figure \ref{Pdr_vs_vdr}  we show that the dependence of $P_{dr}$ vs $\nu_{dr}$ can be  fitted by 
power law $P_{dr} \propto \nu_{dr}^{-1.8\pm0.16} $.  Namely the driving oscillation power $P_{dr}$ decreases when the source (Cyg X-1)  goes to softer states. 
%This is an observational fact.  
Presumably the decay of $P_{dr}$ with 
$\nu_{dr}$ is  also related to the  contraction of Compton cloud.  
The driving oscillations can result from the Rayleigh-Taylor (RT) local instability [see e.g. \citet{c61} and  \citet{t03}] ].
The decay of $P_{dr}$  can be considered as a cumulative effect of the local Rayleigh-Taylor (RT)  instability  
when  the effective area of a given configuration (CC)  undergoing  RT oscillations contracts.  

In Figure \ref{p_x_vs_vdr} we present a comparison of the observable PDS integrated power 
$P_{x}$  (black filled circle) with the model predicted $P_{x,diff}$ (crosses) 
(see Eq. \ref{theory_obs_form}). One can see that the dependence $P_{x,diff}$ on  $\nu_{dr}$ is similar to the observable  
correlation $P_{x}$ vs $\nu_{dr}$.  

Note that we obtain the factor $C\sim 4$ (see Eq. \ref{theory_obs_form}) 
by shifting a set of the values of  $P_{x,diff}$  along  Y-axis to fall on top of $P_{x}$ values. 
The power-law  $P_{x} \propto \nu_{dr}^{-0.48\pm 0.03}$ fits
 the dependence of the theoretical and observable integrated powers  vs the 
driving oscillation frequency.

\section{The Reynolds number  of the accretion flow in Compton cloud configuration}

\cite{tlm98}, hereafter TLM98  introduced the Reynolds number   
$\rm Re= V_{R}R/\hat\nu$  ($\gamma$ in their notation)  where $V_{R}$,  $\hat\nu$  are an average radial velocity,  an average  viscosity  over a given configuration respectively  and $R$ is a configuration scale.   They demonstrate that the size of the transition layer (CC)  between the fast rotating accretion disk and the relatively slow rotating central object (either BH or NS)  strongly depends on the Reynolds number $\rm Re$.  It is worth noting that  $\alpha-$viscosity parameter   introduced by \cite{ss73} is  related to the TL (CC)  parameter $\rm Re$.  Given that $\alpha\sim \hat \nu/(V_{s}H)$, where $V_s$ is the sound velocity in the ``disk-like'' configuration and $H$ is a half of the configuration vertical size, we obtain that
\begin{equation}
\rm Re\sim \lambda/\alpha. 
\label{alpha_re}
\end{equation}
To infer this  formula one should  assume that $\lambda=H/R$ and $V_s\sim V_R$. 

TSA07 has already shown that there is a possibility to infer $\rm Re$, and consequently $\alpha-$parameter 
(see Eq. \ref{alpha_re}),  using the RXTE  observations of BHs and NSs. 
This determination can be done  if   a particular RXTE  power spectrum (PDS) has   the white-red noise  (WRN)  component and also if the QPO low frequency $\nu_L$  is present there.  TSA07 argue that  the relation between observed frequencies $\nu_L$  and $1/t_0$ seen in Cyg X-1 PDSs is similar to the theoretical relation between  the diffusion frequency $\nu_{diff}$ and the frequency of CC volume (magneto-acoustic) oscillations $\nu_{MA}$ [see e.g. \cite{to99}]. 
This fact leads them to conclude that $\nu_L$ can be considered as the frequency of  the magneto-acoustic oscillations which theory was developed by \cite{tbw01}, hereafter TBW01.

%because the presence of magnetic field in the disk can be excluded.}  

\begin{figure}
%\begin{figure}[ptbptbptb]
\includegraphics[width=3.8in,height=5.7in,angle=-90]{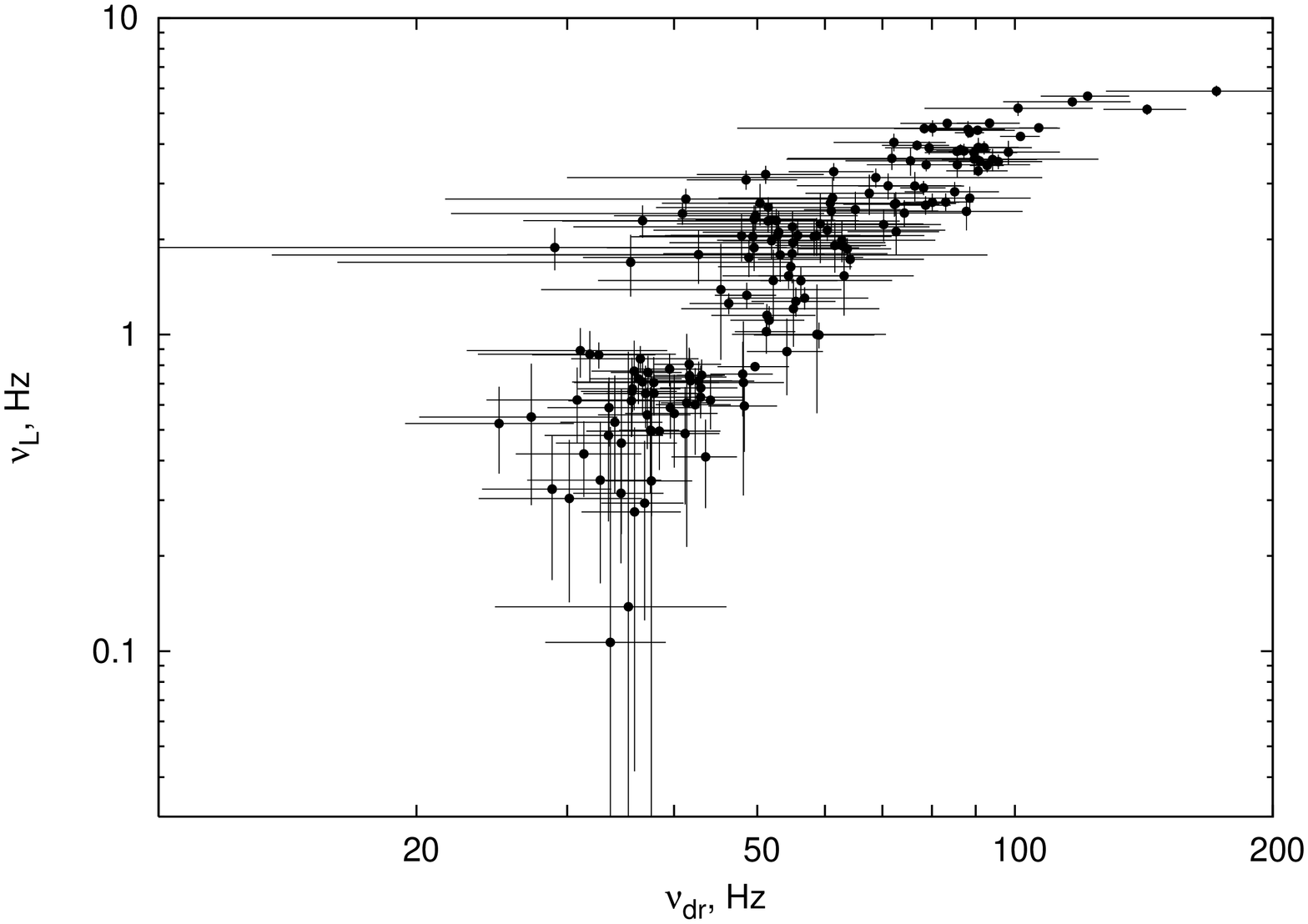}
\includegraphics[width=3.8in,height=5.7in,angle=-90]{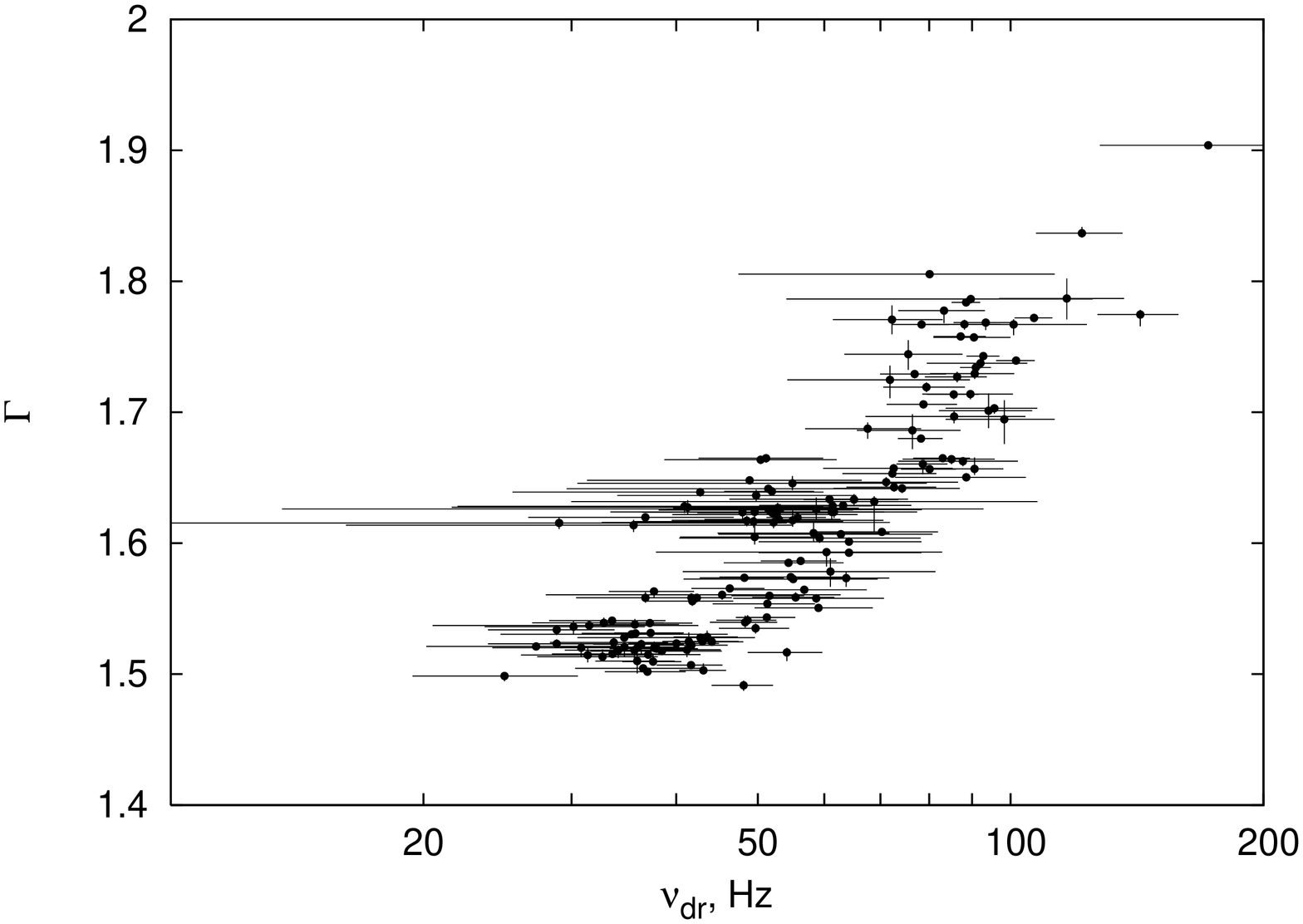}
\caption{Upper panel: low QPO frequency $\nu_L$ vs driving QPO frequency  $\nu_{dr}$
lower panel: photon index $\Gamma$ vs  $\nu_{dr}$ .
 }
\label{nu_dr_vs_gamma}
\end{figure}

\begin{figure}[ptbptbptb]
\includegraphics[width=5in,height=7.1in,angle=-90]{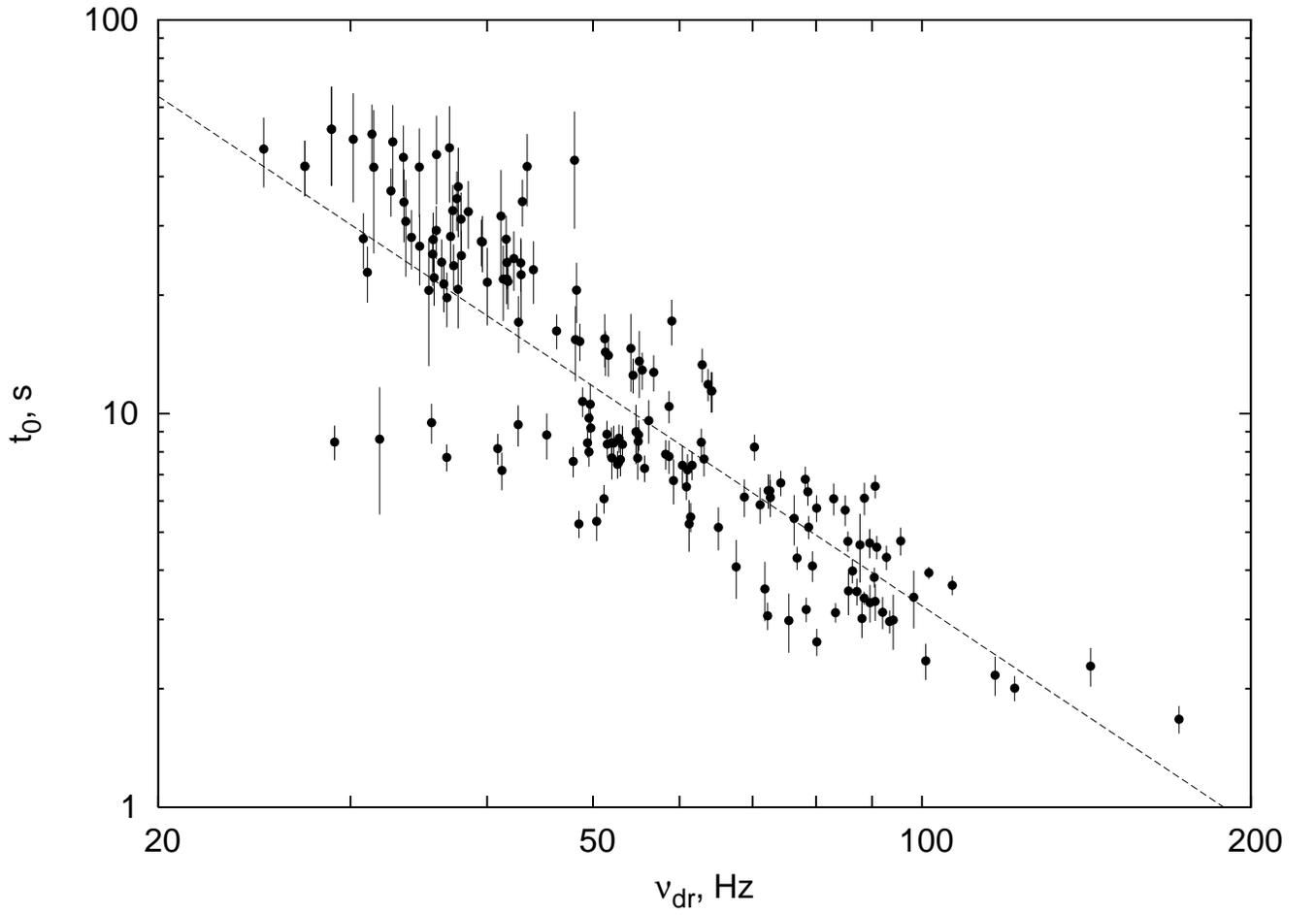}
\caption{The best-fit model parameter, diffusion time scale $t_0$ vs $\nu_{dr} $. The dashed line is the best fit power law $t_0 \propto \nu_{dr}^{-2.13\pm0.14} $. 
 }
\label{t0_vs_vdr}
\end{figure}

\begin{figure}[ptbptbptb]
\includegraphics[width=5in,height=7.1in,angle=-90]{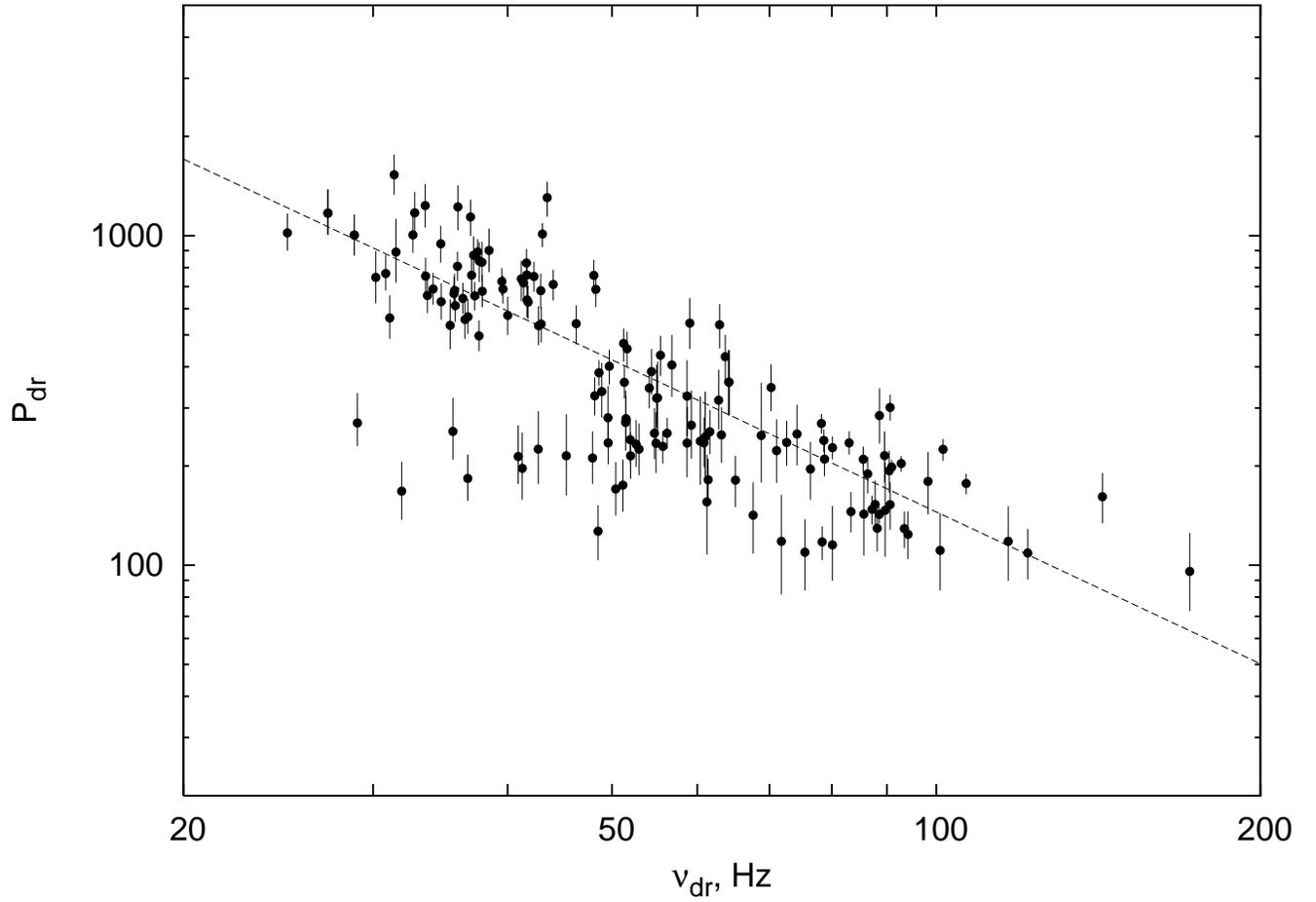}
\caption{Inferred $P_{dr} $ vs $\nu_{dr} $. The dashed line is the best fit power law $P_{dr} \propto \nu_{dr}^{-1.8\pm 0.16} $. 
 }
\label{Pdr_vs_vdr}
\end{figure}

\begin{figure}[ptbptbptb]
\includegraphics[width=5in,height=7.1in,angle=-90]{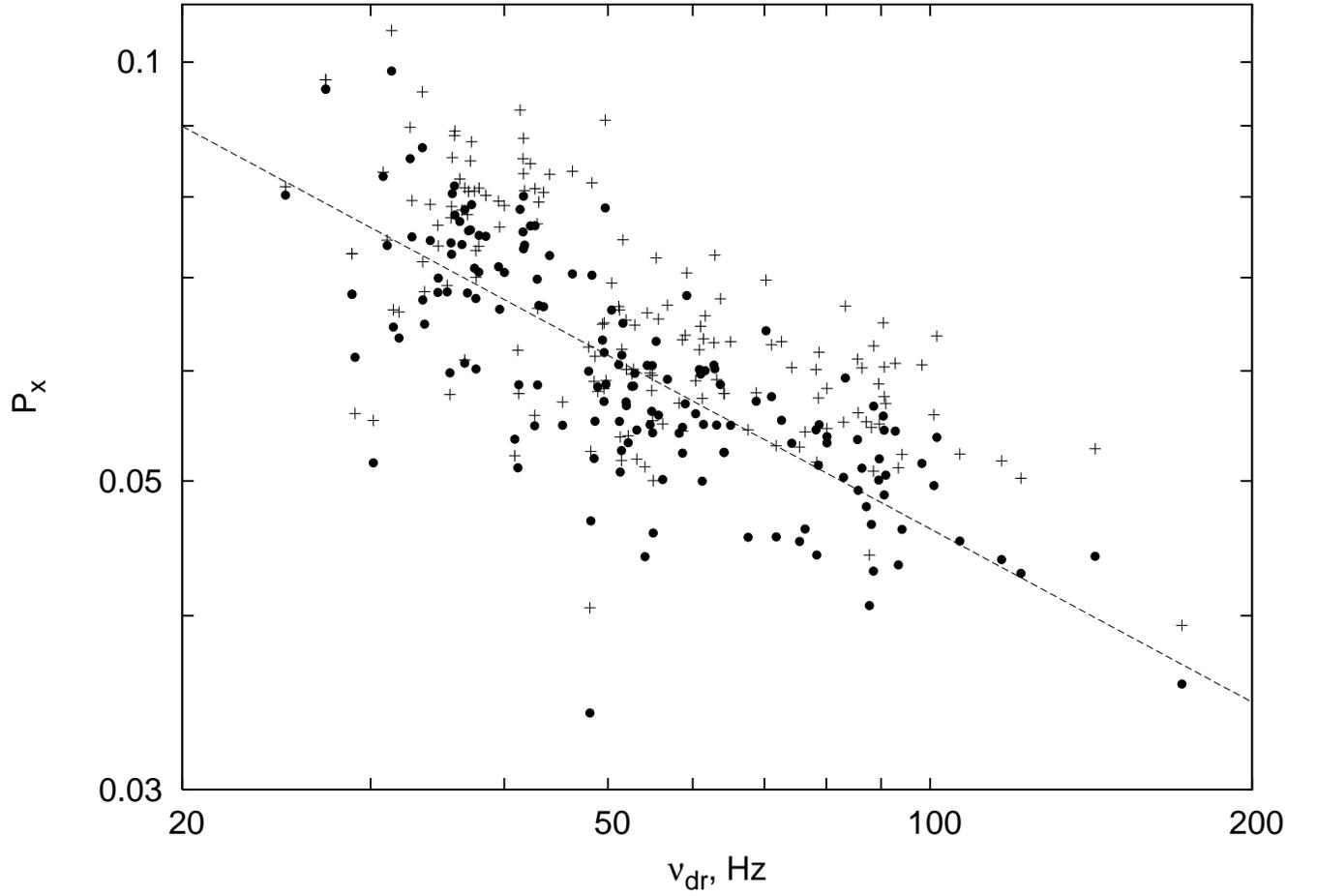}
\caption{ Comparison of the model $P_{x,diff}$ vs $\nu_{dr}$ (crosses)  (using  Eq. \ref{theory_obs_form}) with 
the observable  $P_{x}$ vs $\nu_{dr}$ (black filled circle) 
The dashed line is the best-fit power law $P_{x} \propto \nu_{dr}^{-0.48\pm 0.03} $.
 }
\label{p_x_vs_vdr}
\end{figure}

\begin{figure}[ptbptbptb]
\includegraphics[width=5in,height=7.1in,angle=-90]{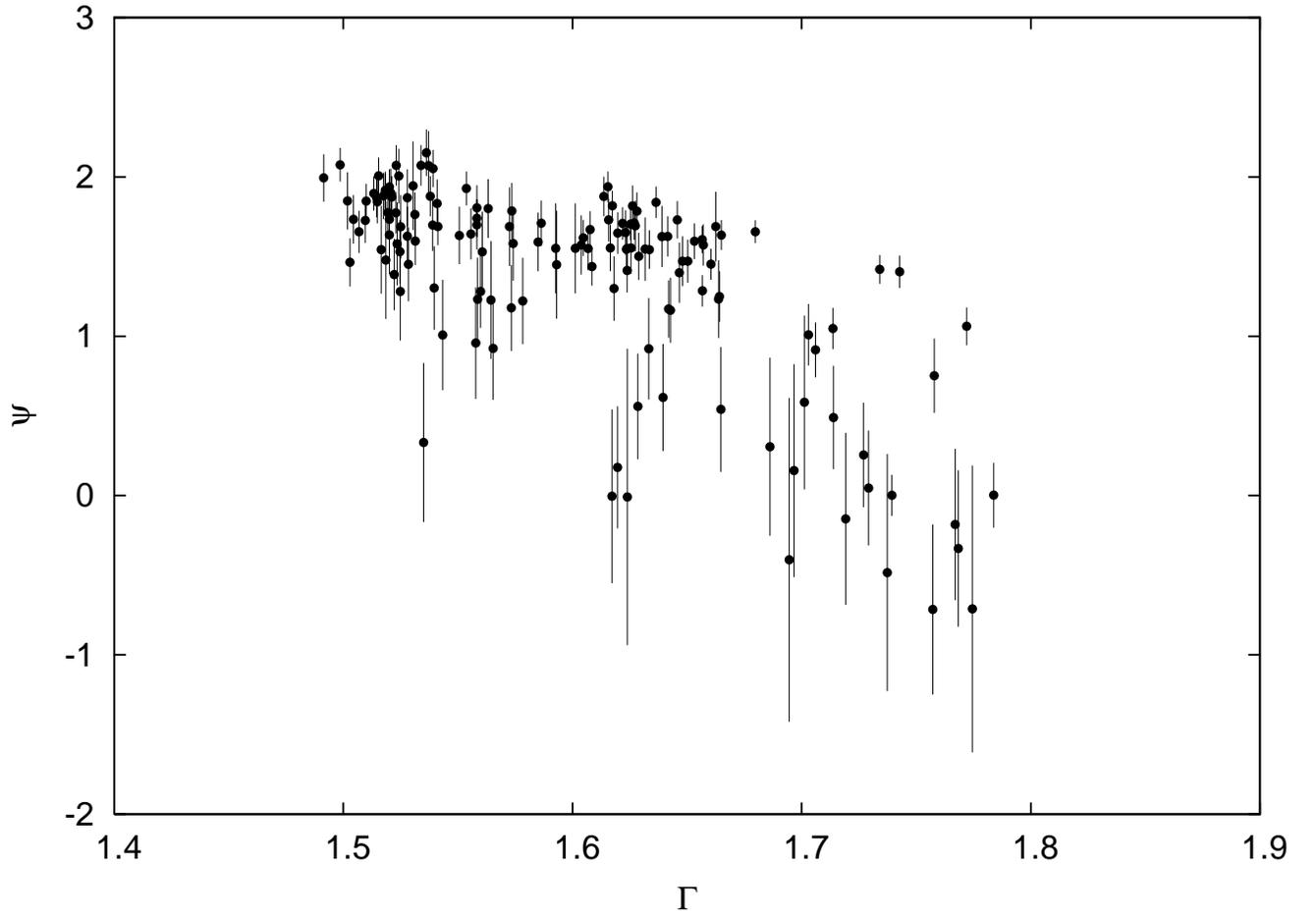}
\caption{The best-fit index of the viscosity distribution $\psi$  vs 
$\Gamma$. 
 }
\label{psi_wrn}
\end{figure}

Using the best-fit parameters of the PDSs we can even infer the evolution of  
%physical parameters   of the source such as
% magneto-acoustic QPO frequency $\nu_{MA}$ and 
the Reynolds number of the accretion flow  in the Compton cloud (CC),  with the change of  photon index $\Gamma$.
%There is only difference that in the intermediate state we see evidence of the presense 
%of two extended configurations
%as  sources of fluctuations, as in low/hard and high/soft states the  band-limited noise and the broad frequency band (``red'') 
%noise are dominated in the PDSs  respectively.
In fact, TSA07  relate   diffusion time $t_0$  with ${\rm Re}$ and  a magneto-acoustic QPO frequency $\nu_{MA}$ as
%(see Eq. \ref{t0})
\begin{equation}
t_0={4\over{3}}{4\over{(4-\psi)^2}}\left[{{V_{R}R_0}\over{\hat\nu(R_0)}}\right]\left(\frac{R_0}{V_{R}}\right)= 
{4\over{3}}{4\over{(4-\psi)^2}}\frac{\rm Re}{a_{MA}\nu_{MA}},
\label{t0_mod}
\end{equation}
where $V_R\sim V_s$ and $a_{MA}$ is a numerical coefficient. 
%To relate $V_{MA}/R_0$ ratio with $\nu_{MA}$ we use a formula for magneto-acoustic oscillation frequency derived by Titarchuk, Bradshaw \& Wood (2001), 
%hereafter TBW01  (see Eqs. 13, 16 and 17 there):
%\begin{equation}
%\nu_{MA}= V_{MA}/(a_{MA}R_0)
%\label{nu_MA}
%\end{equation}
%where $a_{AM}\sim2\pi$ is for a pure acoustic case without magnetic field ($\alpha=0$ in Eqs. 13,  17 in TBW01) 
 %and $a_{AM}\sim 1$ is for a pure magnetic case ($\alpha=6$  see  Eqs. 13, 16 in TBW01).
%The values of $a_{AM}$ presented here are for the free boundary conditions (see TBW01 for details) which are presumably appropriate for the disk-like 
%configurations around BHs.  
Formula (\ref{t0_mod}) leads to the equation
\begin{equation}
{\rm Re}=2\pi \left(\frac{a_{MA}}{2\pi}\right)\frac{3}{4}\frac{(4-\psi)^2}{4}(\nu_L t_0)
\label{Re}
\end{equation}
that allows us to infer a value of $\rm Re$ using the best-fit model parameters $t_0$ and the QPO low frequency  $\nu_L$ 
which is presumably close   to $\nu_{MA}$. Ultimately we can find the evolution of ${\rm Re}$ with the photon index $\Gamma$ given that  $\nu_L$,  $t_0$ and  the viscosity index $\psi$ evolve with $\Gamma$ 
 (see Figs. \ref{nu_dr_vs_gamma}-\ref{t0_vs_vdr} and \ref{psi_wrn}).

\begin{figure}[ptbptbptb]
\includegraphics[width=5in,height=7.1in,angle=-90]{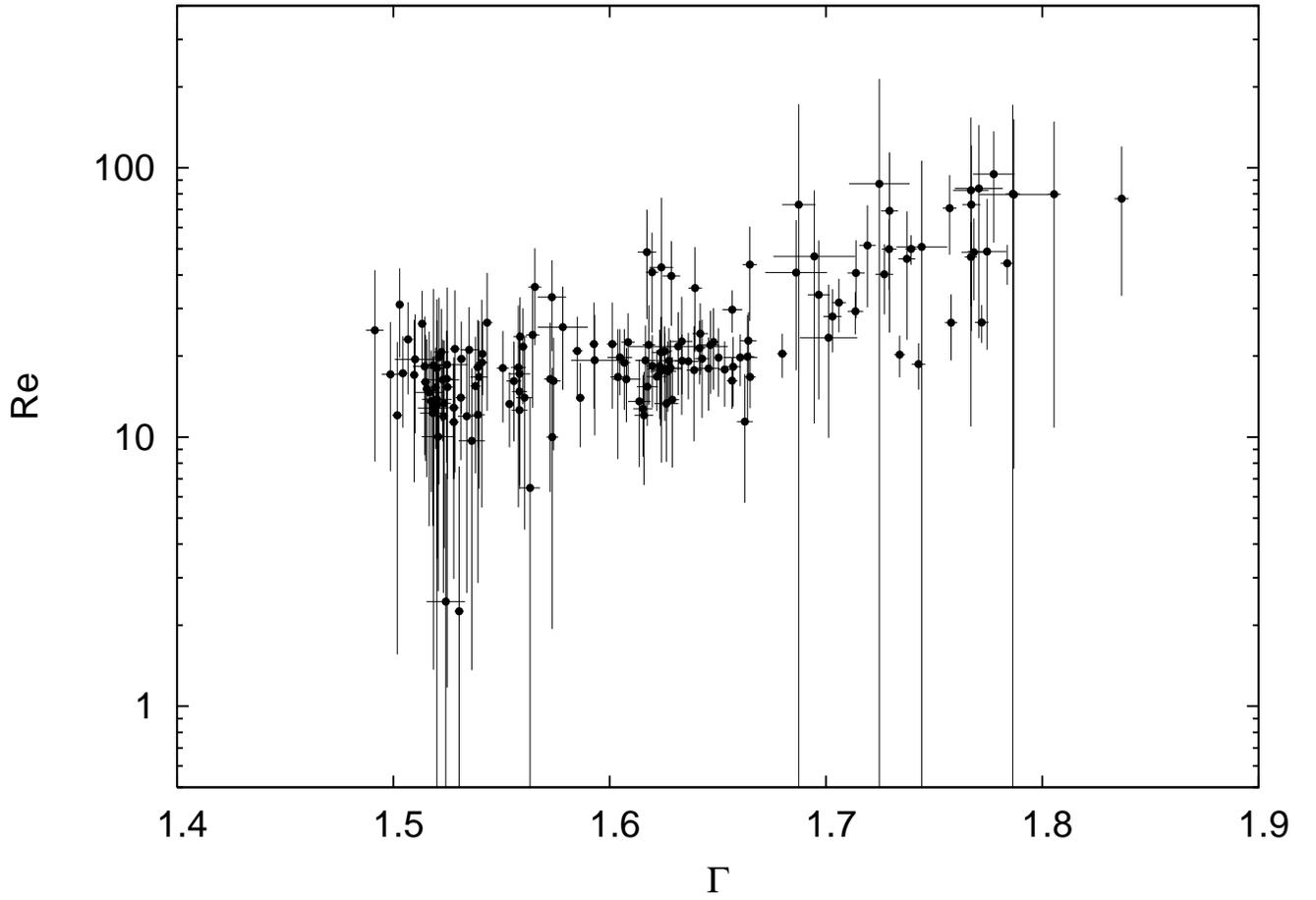}
\caption{Inferred Reynolds number ${\rm Re}$ (using $t_0$, $\nu_L$, $\psi$, and  Eq. \ref{Re})
 vs $\Gamma$. 
 }
\label{re_vs_alpha}
\end{figure}

In Figure \ref{re_vs_alpha} we present the inferred  Reynolds number as a function of the photon index $\Gamma$.
We  use  Eq. (\ref{Re}) where we set $a_{MA}=2\pi$ (see details of this determination of $a_{AM}$  in TBW01 and TSA07) and the observable correlations of $\nu_L$ and $t_0$  with $\Gamma$ (see Figs. \ref{nu_dr_vs_gamma} and \ref{t0_vs_vdr}).   One can see that ${\rm Re}-$number
 steadily increases from 10 to  70 when the source evolves from low/hard state to high/soft state. 
In contrast,  TSA07  found  that ${\rm Re} \sim 8\pm 2.5$. We note, however, that in TSA07 only WRN model was used 
without accounting for driving oscillation distribution, which significantly affects the resulting value for $\psi$.
They also used a limited set of data.

Note the observed behavior of the Re-number  vs $\Gamma$ and mass accretion rate was 
predicted by TLM98,
%Titarchuk, Lapidus \& Muslimov (1998), 
%hereafter TLM98, 
where  they  formulated a transition layer model 
(TLM) and  studied its consequences for observations. It is important to emphasize that the ${\rm Re}-$number  
along with the photon index $\Gamma$, the low frequency QPO $\nu_L$ and the driving frequency $\nu_{dr}$ can be 
considered as characteristics  of the spectral state. All of them correlate with each other.

\section{Photon index-QPO frequency correlation} 

TLM98 showed that the outer (adjustment) radius of the transition layer (CC) $R_{out}$ measured in the dimensionless 
units  with respect to  Schwarzschild radius $R_{\rm S}=2GM/c^2$, 
$r_{out}=R_{out}/R_{\rm S}$, anticorrelates with  ${\rm Re}-$number or photon index $\Gamma$  (spectral state) only.
Thus  $\nu_L$ (or $\nu_{MA}$) as a  ratio $\sim V_{MA}/R_{out}$ should correlate with $\Gamma$ 
(or ${\rm Re}-$number)  
where values $\nu_L$  related to the same $\Gamma$  for different sources should be inversely proportional to a 
mass of the central object (black hole or neutron star). Note that a plasma velocity $V_{MA}$  is also a function 
of $\Gamma$  only.  The comparison of the observed  index--QPO frequency correlations for two 
different sources (with two different masses) should lead to determination of their relative masses with respect each other. 
This is the main idea behind  the method of weighing black holes (TLM98)  recently  applied for BH mass determination in 
 a number of Galactic and extragalactic sources [see TF04; \citet{ft04}; \citet{dtg06};  \citet{str07} and \citet{st07}]. 
 
\section{Discussion and Conclusions} 

 \cite{ct95} proposed  a model  of
the truncated Keplerian  disk  and hot corona located in the innermost part of the source using  arguments in terms of radiation hydrodynamics.
\cite{cgr99},  hereafter CGR99, were the first who supported   this model 
using the specific form of the power spectra of X-ray radiation. 
Particularly they noted that   two-component X-ray spectra (soft multicolour black-body and harder power law) are frequently observed from accreting black holes. These components are presumably associated with the different parts of the accretion flow (optically thick and optically thin respectively) in the vicinity of the compact source. 
They also emphasized that for Cygnus X-1, the overall shape of the power density spectra  in the soft and hard spectral states can be  {\it qualitatively}  explained if the geometrically thin disc is sandwiched by the geometrically thick corona extending in a radial direction up to a large distance from the compact object. In the hard state the thin disc is truncated at some distance from the black hole followed by the geometrically thick flow. The break in the PDS is then associated with the characteristic frequencies in the accretion flow at the thin disc truncation radius.

However there is a difference between CGR99's suggestions and the results of our paper. 
We come to the model of the truncated disk plus corona  using the fit of the theoretical model of the power spectrum (see TSA07) to the data. Whereas CGR99 proposed  a  {\it qualitative} explanation of the data using  Lyubarskii's idea on the formation of the power spectrum in X-ray binaries 
[see L97]. 

Our results confirm the CGR99 conclusion that   a stable optically thick (but geometrically thin)  disk extends down to  small radii  (as indicated by the strong and stable soft component) and whereas prominent variations of the harder component (presumably related to Compton cloud) are present in the broad range  of time-scales (up to at least $10^2$  s;  see Figs. 2-3  in TSA07). We also   confirm   the observed  emergent
spectrum consisting of two components (soft stable component
due to the disc emission and harder highly variable component due to
Comptonization in the corona). The relative contribution of these
two components to the luminosity of an averaged spectrum would
then reflect the ratio of the energy releases in the disc and corona
(or mass accretion rates).  It is also worth noting that our presented analysis indicates  that the high-frequency part of the PDS (QPO and  turnover)  at about 10 Hz and  higher is possibly related  to the  local instabilities operating in the Comptonization region of the energy release
(see CGR99).  
Note that \cite{gz05}  also pointed out the evolution of PDSs in XTE J15550-564 and XTE J1650-500. In particular  they showed a variable power in the Comptonized component  (the PDS high-frequency component) when  a given source evolves  from hard to soft states.  

However there is a difference in details between the CGR99 and results of our papers (see also TSA07).
For example the break in the PDS  frequency range  between 0.01 and 1 Hz is associated with the  characteristic diffusion frequencies of the Compton cloud but not as that in CGR99 ``with the characteristic frequencies in the accretion flow at the thin disc truncation radius.''

\cite{abl05},  hereafter ABL05, suggested  a  model  of the PDS of Cyg X-1 
which successfully presented a natural transition from hard
state through intermediate state to soft state and back, and also allowed them 
to study the behavior of the Lorentzian components in detail. Using this  model
consisting of a power-law and two Lorentzian profiles,  ABL05 were
able to fit the Cyg X-1 PDSs, and to follow the components
from hard state through the transitions and back. By choosing
this  simple  empirical approach for the PDS data analysis ABS05 could
model the major features in all states with only three components. 
The parameters used in ABL05 were the width  and peak frequency
 of the Lorentzians, along with the power at the peak frequency,
The parameter evolution and its correlation between each other can all be described
by continuous functions.

However ABL05 did not address to the nature  of the  components and  continuum of the power spectrum. Their PDS description was pure phenomenological (compare with that in TSA07 and this paper).
    
ABL05 also showed that the PDS
of Cyg X-1 is dominated by the same two Lorentzian components
at all times. They came  to this conclusion based  on the fact that
 the relation of the peak frequencies of these  Lorentzian components follows the pattern seen in both
other BH and NS systems, and  therefore  they proposed that the
physical processes responsible for them cannot be explained by
invoking magnetic fields, a solid surface, or an event horizon.

It is worth noting that we also find  that the diffusion (break) and QPO frequency follows the BH-NS pattern (see details in ST06 and TSA07).   In addition to this  we demonstrate  that  there is a certain correlation between QPO low frequency and driving (high) frequency and photon index (see Fig. 2)  and show a correlation between the diffusion time scale (an inverse of the diffusion frequency) and the driving frequency (see Fig. 3).  Furthermore   we  explain the nature of the observational appearances of the power and photon spectra evolution by  the evolution  of the truncated disk and  corona configurations when a X-ray source evolves from hard state to soft states. 
 We come to the conclusion that the  truncated disk and  corona scenario should  be a common model in BH and NS systems. 

In fact, TLM98   showed that the bounded configuration (Compton cloud) surrounding compact objects
is the transition layer (TL) that is formed as a result of dynamical adjustments of a Keplerian disk to the innermost sub- Keplerian boundary conditions. They argued that this type of adjustment is a generic feature of the Keplerian flow in the presence of the sub-Keplerian boundary conditions near the central object and that it does not necessarily require the presence or absence of a hard surface. TLM98 concluded that
an isothermal sub-Keplerian TL between the NS surface and its last Keplerian orbit forms as a result of this adjustment. The TL model is general and is applicable to both NS and BH systems.

As the conclusions we want to single out

1. That we explain the decay of the emergent time variability of X-ray emission in  compact  sources when these sources evolve from 
low/hard   to high/soft states. We find that the resulting power $P_x$ from Cyg X-1  decays with the driving oscillation 
frequency $\nu_{dr}$ as   $P_x\propto \nu_{dr}^{-0.5}$. 

2. We show that the reciprocal  of the diffusion time scale of the perturbation $t_0^{-1}$,  the low frequency QPO $\nu_L$,
 the driving oscillation 
frequency $\nu_{dr}$, inferred by fitting the  Cyg X-1 PDSs with our diffusion model, increase when the source 
evolves from low/hard  state to  high/soft state.  This behavior of the PDS characteristics implies that the Compton cloud 
contracts towards softer spectral states.  
The  driving oscillations are  probably caused by  the local Rayleigh-Taylor  instability 
which cumulative  $P_{dr}$ decreases when the effective area of the configuration producing the 
RT oscillations  contracts. The decay in driving power leads to the decay in the total observed variability 
power from the source.
Using  the fact that $t_0^{-1}$,  $\nu_L$, $\nu_{dr}$ increase with $\Gamma$ and $P_{dr}$, $P_x$ decrease with 
$\Gamma$  we conclude, as a result of our  analysis, that  the Compton Corona shrinks
when Cyg X-1 goes from low/hard  state to high/soft state. 

3. Our extensive data analysis of the power spectra from Cyg X-1 with an application of the method of 
 ${\rm Re}-$number  determination developed in TSA07 
 indicates  that ${\rm Re}$ related to Compton cloud configuration increases 
 %when the source evolves toward softer states. Our analysis gives us that ${\rm Re}$ increases 
 from values about  10 in low/hard state to that about  70  in high/soft state. 
We confirm the predictions by TLM98 that ${\rm Re}-$number  should increase with index 
and QPO frequencies.  Thus one can conclude that the observable index-QPO correlation is probably driven 
by the increase of ${\rm Re}-$number   when the source evolves from low/hard state to high/soft state. 
It is worth noting that inverse proportionality of the low-frequency QPO with respect   BH mass in the index-QPO 
correlation leads to  the  method weighing BHs  employing  this index-QPO correlation. 

We acknowledge very useful comments and suggestions by the referee. 
%Namely we show  that ${\rm Re}$ correlates with the index.  This  correlation imply

\newpage

\newpage

%\newpage
%\begin{figure}[ptbptbptb]
%\includegraphics[width=5in,height=7.1in,angle=-90]{nul_wrn.ps}
%\includegraphics[scale=1.2,angle=-90]{f2.eps}
%\caption{The best-fit model paramer $\nu_{dr}$ vs QPO low frequency $\nu_L $.
% }
%\label{nul_wrn}
%\end{figure}

\newpage
\newpage

\newpage

\newpage

\newpage

\end{document}